\documentclass[aps, prd, amsmath, floats, floatfix, twocolumn, nofootinbib, showkeys, superscriptaddress]{revtex4-1}
\usepackage{graphicx}
\usepackage{color}
\usepackage{latexsym}
\usepackage[colorlinks]{hyperref}

\newcommand{\be}{\begin{eqnarray}}
\newcommand{\ee}{\end{eqnarray}}

\begin{document}

\title{{\sc relxill\_nk}: a relativistic reflection model for testing Einstein's gravity}
\altaffiliation{Talk given at the {\it International Conference on Quantum Gravity} (26-28 March 2018, Shenzhen, China)}

\author{Cosimo~Bambi}
\email[Speaker and corresponding author: ]{bambi@fudan.edu.cn}
\affiliation{Center for Field Theory and Particle Physics and Department of Physics, Fudan University, 200438 Shanghai, China}
\affiliation{Theoretical Astrophysics, Eberhard-Karls Universit\"at T\"ubingen, 72076 T\"ubingen, Germany}

\author{Askar~B.~Abdikamalov}
\affiliation{Center for Field Theory and Particle Physics and Department of Physics, Fudan University, 200438 Shanghai, China}

\author{Dimitry~Ayzenberg}
\affiliation{Center for Field Theory and Particle Physics and Department of Physics, Fudan University, 200438 Shanghai, China}

\author{Zheng~Cao}
\affiliation{Center for Field Theory and Particle Physics and Department of Physics, Fudan University, 200438 Shanghai, China}

\author{Honghui~Liu}
\affiliation{Center for Field Theory and Particle Physics and Department of Physics, Fudan University, 200438 Shanghai, China}

\author{Sourabh~Nampalliwar}
\affiliation{Theoretical Astrophysics, Eberhard-Karls Universit\"at T\"ubingen, 72076 T\"ubingen, Germany}

\author{Ashutosh~Tripathi}
\affiliation{Center for Field Theory and Particle Physics and Department of Physics, Fudan University, 200438 Shanghai, China}

\author{Jingyi~Wang-Ji}
\affiliation{Center for Field Theory and Particle Physics and Department of Physics, Fudan University, 200438 Shanghai, China}
\affiliation{MIT Kavli Institute for Astrophysics and Space Research, MIT, Cambridge, MA 02139, USA}

\author{Yerong~Xu}
\affiliation{Center for Field Theory and Particle Physics and Department of Physics, Fudan University, 200438 Shanghai, China}

\date{\today}

\begin{abstract}
Einstein's theory of general relativity was proposed over 100~years ago and has successfully passed a large number of observational tests in the weak field regime. However, the strong field regime is largely unexplored, and there are many modified and alternative theories that have the same predictions as Einstein's gravity for weak fields and present deviations when gravity becomes strong. {\sc relxill\_nk} is the first relativistic reflection model for probing the spacetime metric in the vicinity of astrophysical black holes and testing Einstein's gravity in the strong field regime. Here we present our current constraints on possible deviations from Einstein's gravity obtained from the black holes in 1H0707--495, Ark~564, GX~339--4, and GS~1354--645.
\end{abstract}

\keywords{General relativity; Black boles; X-ray astronomy}

\maketitle


\section{Introduction}

Einstein's theory of gravity, the theory of general relativity, has undergone a variety of observational tests since it was first proposed a century ago, primarily with experiments in the Solar System and radio observations of binary pulsars~\cite{will}. While it has largely been successful, questions remain on its validity in the so called strong field regime. There are indeed many modified and alternative theories that have the same predictions as Einstein's gravity for weak fields and present deviations when gravity becomes strong. Testing strong gravity is among the priorities in contemporary physics.

Astrophysical black holes are the most extreme objects that can be found in the Universe and ideal laboratories for testing strong gravity~\cite{r-em,r-em2,r-gw}. In 4-dimensional Einstein's gravity, the only stationary and asymptotically flat, vacuum black hole solution, which is regular on and outside of the event horizon, is the Kerr metric~\cite{kerr}. Kerr black holes are relatively simple objects, and are completely described by only two parameters, the mass $M$ and the spin angular momentum $J$ of the black hole. This is the result of the celebrated ``no-hair theorems''~\cite{h1,h2,h3}.

It is remarkable that the spacetime metric around an astrophysical black hole formed from gravitational collapse should be well approximated by the stationary Kerr solution of Einstein's gravity. After the gravitational collapse of the progenitor body and the formation of the black hole, the spacetime quickly moves to the Kerr solution with the emission of gravitational waves~\cite{k1,k2} (``black holes rapidly go bald''). Astrophysical bodies can have a non-vanishing electric charge. However, the equilibrium electric charge for a macroscopic black hole is very small and completely negligible for the spacetime metric~\cite{k3,book}. The presence of an accretion disk around the black hole, or of a companion star orbiting the black hole, has a very small impact on the background metric in the strong gravity regions and can be safely ignored~\cite{k4,k5}. For example, the ratio between the mass of the accretion disk and the mass of the black hole in an X-ray binary system can be $m/M \sim 10^{-10}$. This is a small quantity that can be used as an expansion parameter to take into account the correction on the background metric and, even assuming that the mass of the accretion disk were at a point close to the black hole, the correction to the Kerr metric would be of the order of $m/M$. If we consider the companion star, the expansion parameter would be $M_{\rm c}/r < 10^{-6}$, where $M_{\rm c}$ and $r$ are, respectively, the mass of the companion star and the distance from the companion star. Even in this case, we can expect corrections to the Kerr metric of the order of $M_{\rm c}/r$ and are negligible. In the end, macroscopic deviations for the Kerr solution may be possible in the presence of new physics, such as modified classical gravity~\cite{new1}, quantum gravity effects~\cite{new2,new3,new4}, or exotic matter~\cite{new5,new6}.

From astrophysical observations, we know two main types of astrophysical black holes. Stellar-mass black holes ($M \approx 3$-100~$M_\odot$) are the natural product of the gravitational collapse of heavy stars, after the latter have exhausted all their nuclear fuel~\cite{bh1}. Supermassive black holes ($M \sim 10^5$-$10^{10}$~$M_\odot$) are found at the center of many galaxies~\cite{bh2}. All these objects are thought to be black holes because this is the most natural interpretation in the framework of conventional physics. For example, in the case of stellar-mass black holes in X-ray binary systems, we can get a dynamical measurement of the mass of the black hole. If the latter exceeds 3~$M_\odot$, which is the maximum mass for a neutron star~\cite{max1,max2}, and we can exclude it is a normal star, the most natural explanation is that it is a black hole. In the case of the supermassive black hole at the center of the Galaxy, we can argue it is too massive, compact, and old to be a cluster of non-luminous bodies like neutron stars~\cite{maoz}. The non-detection of electromagnetic radiation from the putative surface of all these objects is also consistent with the conjecture that they are instead a black hole with an event horizon~\cite{event1,event2}. The recent detections of gravitational waves are consistent with the signals expected from black holes in general relativity~\cite{ligo1,ligo2}, even if current data cannot put strong constraints on alternative scenarios~\cite{ligo3,ligo4}. Despite this body of evidence, we still do not know if the spacetime metric around these objects is described by the Kerr solution, as would be required in Einstein's gravity.


\section{X-ray reflection spectroscopy}

\begin{figure}[t]
\begin{center}
\includegraphics[type=pdf,ext=.pdf,read=.pdf,width=8.5cm]{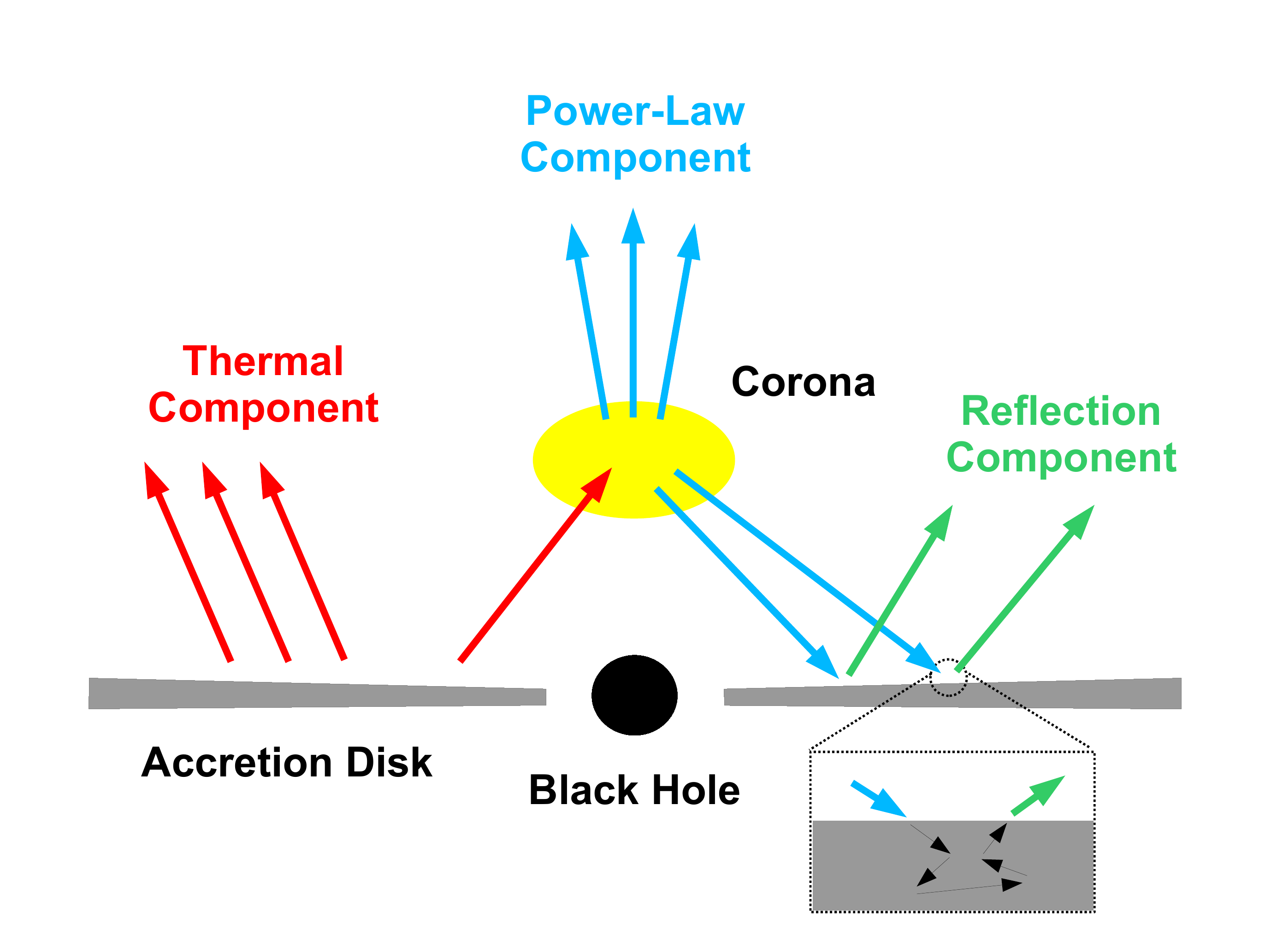}
\end{center}
\vspace{-0.5cm}
\caption{Disk-corona model. \label{f-corona}}
\end{figure}

The standard approach to analyze black holes in an astrophysical setting is the disk-corona model~\cite{book,ann}, where black holes are surrounded by a geometrically thin disk and possess a ``corona'', as shown in Fig.~\ref{f-corona}. In the case of a stellar-mass black hole in a binary system, the material of the disk comes from the companion star. In the case of a supermassive black hole in a galactic nucleus, the disk is created by the material of the interstellar medium around the object. The disk emits like a blackbody locally, and as a multi-temperature blackbody when integrated radially (this is known as the {\it thermal component}). The temperature of the accretion disk depends on the black hole mass, the mass accretion rate, and the radial coordinate of the emission point in the accretion disk. For a black hole accreting at about 10\% of its Eddington limit, the thermal spectrum of the inner part of the accretion disk is in the soft X-ray band (0.1-1~keV) for stellar-mass black holes and in the optical/UV band (1-10~eV) for the supermassive ones. The corona is a hotter ($\sim 100$~keV), usually optically thin, cloud near the black hole. For instance, it may be the base of a jet or some atmosphere above the accretion disk, but the exact morphology is not well understood.

Thermal photons from the disk can gain energy via inverse Compton scattering off the hot electrons in the corona, and transform into X-rays with a characteristic {\it power-law component}. These reprocessed photons in turn illuminate the disk, producing a {\it reflection component} with fluorescent emission lines. The most prominent feature of the reflection spectrum is usually the iron K$\alpha$ line, with emission lines at 6.4~keV in the case of neutral or weakly ionized iron and shifts up to 6.97~keV for H-like ions.

While the iron K$\alpha$ line is a narrow line in the rest-frame of the gas, relativistic effects due to the gravity of the central black hole cause this \textit{line} to be broadened and skewed for observers far away. In the presence of high quality data and with the correct astrophysical model, analysis of the reflection spectrum can be a powerful tool for probing the strong gravitational fields of accreting black holes. The method was originally proposed and developed for measuring black hole spins under the assumption that the metric around astrophysical black holes is described by the Kerr solution~\cite{i1,i2}. More recently, the technique has been proposed for testing Einstein's gravity in the strong field regime~\cite{i3,i4,i5,i6,i7,i8,i9,i10,i11,i12,i13,i14,i15,i16}. Note that spin measurements (if we assume the Kerr metric) or tests of the Kerr metric require fitting the whole reflection spectrum, not just the iron line, even if often (but not always) the iron line is the feature that primarily determines the measurement of the parameters of the background metric in the strong gravity region.


\section{The relativistic reflection model {\sc relxill\_nk}}

\subsection{{\sc relxill}}

Currently, the most advanced model for the calculation of the reflection spectrum of the accretion disk around Kerr black holes is {\sc relxill}~\cite{relxill1,relxill2}. {\sc relxill} is the result of the merger of the reflection code {\sc xillver}~\cite{x1,x2} with the relativistic convolution model {\sc relconv}~\cite{relconv}. We can write
\begin{flushleft}
\hspace{0.5cm} {\sc relxill} $\sim$ {\sc relconv} $\times$ {\sc xillver} .
\end{flushleft}

{\sc xillver} provides the reflection model in the rest-frame of the emitting gas at every point of the accretion disk. Compared to all earlier codes, {\sc xillver} has a superior treatment of the radiative transfer and of the ionization balance, by implementing the most complete atomic database for modeling synthetic photoionized X-ray spectra. The microphysics captured by {\sc xillver} is much more rigorous than for any earlier code, principally because of the detailed treatment of the K-shell atomic properties of the prominent ions. Despite that, there are still a number of limitations in this reflection model: the electron density is fixed to $10^{15}$~cm$^{-3}$ and independent of vertical height, with the exception of iron all elemental abundances are assumed to be Solar, thermal photons of the accretion disk are ignored in the radiative transfer calculations, etc.

{\sc relconv} is a convolution model. It requires as input the reflection spectrum in the rest frame of the gas at every point of the accretion disk and provides as output the reflection spectrum of an accretion disk around a Kerr black hole as it would be detected by a faraway observer. The model assumes that the accretion disk is infinitesimally thin and lies on the equatorial plane, perpendicular to the black hole spin. The particles of the gas of the accretion disk are supposed to follow nearly geodesic circular orbits on the equatorial plane. With a similar set-up, {\sc relconv} integrates the reflection spectrum of the disk over all radii and includes the relativistic effects in a Kerr spacetime affecting the photons propagating from the emission point on the disk to the detection point in the flat faraway region.

\subsection{Testing the Kerr black hole hypothesis}

There are two natural approaches to test the spacetime metric around astrophysical black holes. They can be called, respectively, {\it top-down} and {\it bottom-up} methods.

In the top-down approach, we want to test a specific alternative theory of gravity in which uncharged black holes are not described by the Kerr solution. In such a case, we fit the astrophysical data both with a Kerr model and with a model employing the non-Kerr spacetime of the gravity theory under investigation. Depending on the quality of the available data and of the theoretical model, we may be able to check whether the astrophysical observations prefer the Kerr metric of Einstein's gravity or the non-Kerr solution of the alternative theory. There are two problems if we follow this method. Firstly, there are a large number of alternative scenarios to Einstein's gravity, and none seems to be more motivated than the others, so we should repeat this study for every alternative theory of gravity. Secondly, typically we do not know the rotating black hole solutions in alternative theories of gravity. The problem is in solving the corresponding field equations for a stationary and axisymmetric solution. Usually, we know the non-rotating solutions or some approximated solutions in the slow rotation limit, but only in exceptional cases do we know the complete rotating solutions.

In the bottom-up approach, we employ a parametrized metric to describe the background geometry of the spacetime in the astrophysical model. Such a parametrized metric is characterized by the mass $M$ and the spin angular momentum $J$ of the compact object generating the gravitational field, as well as by a number of ``deformation parameters''. If we set all the deformation parameters to zero, we have to recover the Kerr metric of Einstein's gravity. If at least one of the deformation parameters is non-vanishing, we have deviations from the Kerr metric. The strategy is thus to fit the astrophysical data with this parametrized model and try to measure the deformation parameters. If the astrophysical data require vanishing deformation parameters, the Kerr black hole hypothesis is verified. If at least one of the deformation parameters were non-vanishing, astrophysical data would prefer a spacetime metric different from the Kerr solution. In general, the parametrized metric and its deformation parameters do not have a clear physical meaning, because the metric is not a solution of some alternative theory of gravity. Their significance instead lies in the fact that they capture
deviations from a Kerr metric.

\subsection{{\sc relxill\_nk}}

Recently, we expanded the {\sc relxill} model to {\sc relxill\_nk}, which incorporates non-Kerr metrics~\cite{relxillnk}. Since we only want to test the background metric around black holes, assuming that atomic physics is the same, the new model is
\begin{flushleft}
\hspace{0.5cm} {\sc relxill\_nk} $\sim$ {\sc relconv\_nk} $\times$ {\sc xillver} .
\end{flushleft}
{\sc relconv\_nk} is a convolution model for non-Kerr metrics and we do not change the reflection code {\sc xillver}.

{\sc relxill\_nk} is presently the only reflection model for testing the strong gravity region of astrophysical black holes. In the current version, {\sc relxill\_nk} incorporates the non-Kerr metric proposed by Johannsen~\cite{j}. In Boyer-Lindquist coordinates, the line element of the Johannsen metric reads (we use units in which $G_{\rm N} = c = 1$)~\cite{j}
\be
ds^2 &=& -\frac{\tilde{\Sigma}\left(\Delta-a^2A_2^2\sin^2\theta\right)}{B^2}dt^2 \nonumber\\
&&-\frac{2a\left[\left(r^2+a^2\right)A_1A_2-\Delta\right]\tilde{\Sigma}\sin^2\theta}{B^2}dtd\phi \nonumber\\
&&+\frac{\tilde{\Sigma}}{\Delta A_5}dr^2+\tilde{\Sigma}d\theta^2\nonumber\\
&&+\frac{\left[\left(r^2+a^2\right)^2A_1^2-a^2\Delta\sin^2\theta\right]\tilde{\Sigma}\sin^2\theta}{B^2}d\phi^2
\ee
where
\be
&&a=J/M \, , \quad B=\left(r^2+a^2\right)A_1-a^2A_2\sin^2\theta \, , \nonumber\\
&& \tilde{\Sigma}=\Sigma+f \, , \quad \Sigma=r^2+a^2\cos^2\theta \, , \nonumber\\ 
&& \Delta=r^2-2Mr+a^2 \, .
\ee
and the four free functions $f$, $A_1$, $A_2$, and $A_5$ are
\be
f &=& \sum^{\infty}_{n=3}\epsilon_n\frac{M^n}{r^{n-2}} \, , \nonumber\\
A_1 &=& 1 + \sum_{n=3}^{\infty}\alpha_{1n}\left(\frac{M}{r}\right)^n \, , \nonumber\\
A_2 &=& 1 + \sum_{n=2}^{\infty}\alpha_{2n}\left(\frac{M}{r}\right)^n \, , \nonumber\\
A_5 &=& 1 + \sum_{n=2}^{\infty}\alpha_{5n}\left(\frac{M}{r}\right)^n \, .
\ee
The metric elements depend on the mass and spin of the black hole as well as on four free functions that measure potential deviations from the Kerr solution. The first order deformation parameters in these free functions are $\epsilon_3$, $\alpha_{13}$, $\alpha_{22}$, and $\alpha_{52}$. This metric exactly reduces to the Kerr solution for $\epsilon_3=\alpha_{13}=\alpha_{22}=\alpha_{52}=0$.

In the Kerr spacetime, the condition for the existence of an event horizon is $| a_* | \le 1$, where $a_* = a/M = J/M^2$ is the dimensionless spin parameter. For $| a_* | > 1$, there is no horizon, and the singularity at $r=0$ is naked. In the Johannsen spacetime, we still have the condition $| a_* | \le 1$. Moreover, in order to exclude a violation of Lorentzian signature or the existence of closed time-like curves in the exterior region, we have to impose that the metric determinant is always negative and that $g_{\phi\phi}$ is never negative for radii larger than the radius of the event horizon. These conditions lead to the following constraints on the first-order deformation parameters~\cite{j}
\be
\label{eq:boundary}
\alpha_{13}, \epsilon_{3} & \ge & - \left( 1 + \sqrt{1 - a_*^2} \right)^3 \, , \nonumber\\
\alpha_{22}, \alpha_{52} & \ge &- \left(1 + \sqrt{1 - a_*^2} \right)^2 \, .
\ee
In the next section, we will review the constraints on $\alpha_{13}$ obtained so far from the analysis of X-ray data of some sources assuming that all the other deformation parameters vanish. Since the metric is singular for $B = 0$, we impose that $B$ never vanishes for radii larger than the event horizon, and this implies the following constraint on $\alpha_{13}$
\be
\label{eq:boundary-2}
\alpha_{13} \ge - \frac{1}{2} \left( 1 + \sqrt{1 - a_*^2} \right)^4 \, ,
\ee
In the studies reported below, we only consider the parameter space satisfying the constraint in Eq.~(\ref{eq:boundary-2}), which is stronger than the requirement in Eq.~(\ref{eq:boundary}).

\section{Observational constraints}

As of now, we have applied our {\sc relxill\_nk} to a few sources: the supermassive black holes in 1H0707--495~\cite{0707} and in Ark~564~\cite{564}, and the stellar-mass black holes in the X-ray binary systems GX~339--4~\cite{339} and GS~1354--645~\cite{1354}. In what follows, we present the current constraints on the Johannsen deformation parameter $\alpha_{13}$ assuming that all the other deformation parameters vanish. Note that we have only analyzed a few sets of data of these objects. It is likely that a more detailed analysis of all the available X-ray data of these sources can provide stronger constraints on the deformation parameter $\alpha_{13}$.

\subsection{1H0707--495}

1H0707--495 is a Narrow Line Seyfert~1 galaxy. The supermassive black hole in its galactic nucleus looks to be a promising source for testing the Kerr metric using {\sc relxill\_nk}. The spectrum of this source shows significant edge features, which are commonly interpreted as an extremely strong reflection component. Moreover, previous studies that assumed the Kerr metric and a reflection dominated spectrum found that the inner radius of the accretion disk is very close to the compact object, and this increases the relativistic effect in the reflection spectrum.

\begin{figure*}[t]
\begin{center}
\includegraphics[type=pdf,ext=.pdf,read=.pdf,width=8.5cm]{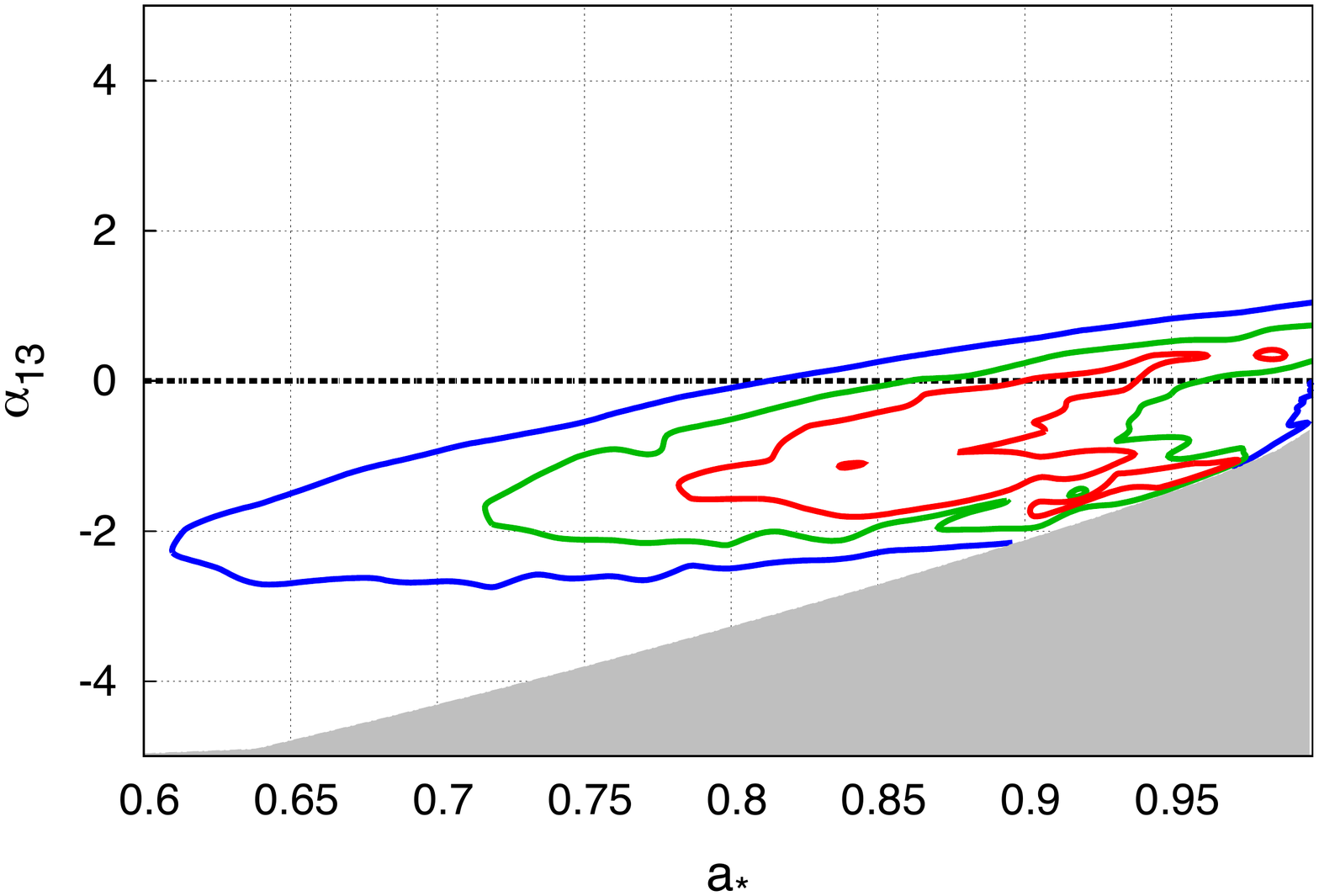}
\hspace{0.5cm}
\includegraphics[type=pdf,ext=.pdf,read=.pdf,width=8.5cm]{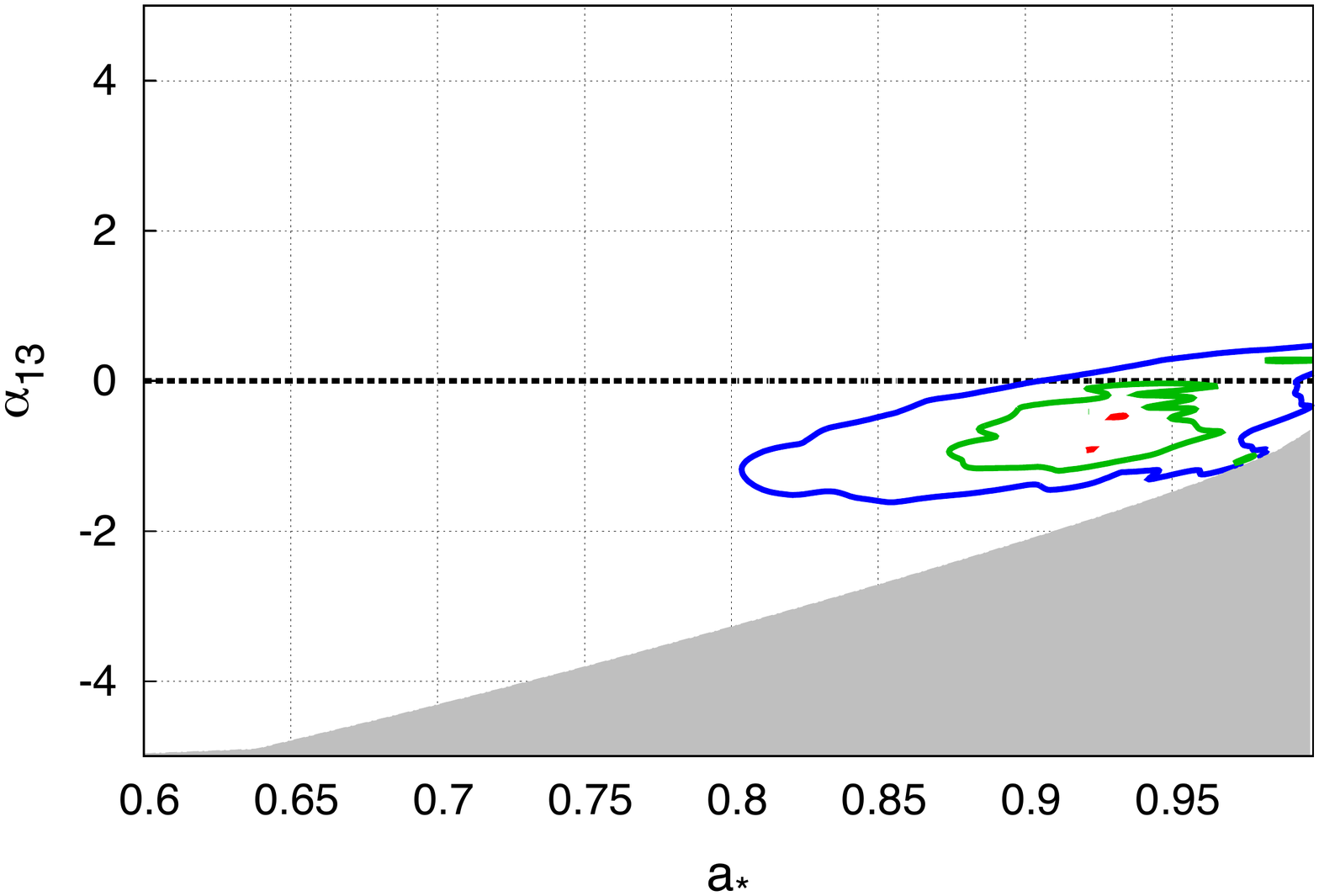}
\end{center}
\vspace{-1.2cm}
\caption{Constraints on the spin parameter $a_*$ and the Johannsen parameter $\alpha_{13}$ from the study in Ref.~\cite{0707} of the 2011 \textit{XMM-Newton} data of 1H0707--495 if we fit the ``soft excess'' around 1~keV with a thermal model (left panel) or if we employ a double reflection model (right panel). The red, green, and blue lines indicate, respectively, the 68\%, 90\%, and 99\% confidence level curves for two relevant parameters. The grayed region is ignored because those spacetimes violate the constraint in Eq.~(\ref{eq:boundary-2}). See the text and Ref.~\cite{0707} for more details. Note that the constraints reported in Ref.~\cite{0707} are slightly different because that study used an earlier version of {\sc relxill\_nk}, and therefore the constraints reported here should be more reliable. \label{f-0707a}}
\end{figure*}

In Ref.~\cite{0707} we analyzed an observation with \textit{XMM-Newton} of 2011 and, separately, three observations with \textit{NuSTAR} (with simultaneous snapshots of \textit{Swift}) of 2014. In the case of the 2011 data, we fit the data with two different models that provide equally good fits. In the first model, we employed the thermal disk model {\sc diskbb} to fit the ``soft excess'' around 1~keV, and the complete model is
\begin{flushleft}
\hspace{0.5cm} {\sc tbabs} $\times$ ({\sc relxill\_nk} $+$ {\sc diskbb}) .
\end{flushleft}
In the second model, we employed a double reflection spectrum and the complete model is
\begin{flushleft}
\hspace{0.5cm} {\sc tbabs} $\times$ ({\sc relxill\_nk} $+$ {\sc relxill\_nk}) .
\end{flushleft}
With the available data, it is not possible to say which one, if any, is correct or wrong, because they both provide good fits. The resulting constraints on the spin parameter $a_*$ and the Johannsen parameter $\alpha_{13}$ are shown in Fig.~\ref{f-0707a}. The red, green, and blue lines indicate, respectively, the 68\%, 90\%, and 99\% confidence level curves for two relevant parameters. The grayed region is ignored because those spacetimes violate the constraint in Eq.~(\ref{eq:boundary-2}). Note that the constraints shown in Fig.~\ref{f-0707a} are slightly different from those reported in Fig.~1 in Ref.~\cite{0707}, because those in Fig.~\ref{f-0707a} have been obtained with a more recent version of {\sc relxill\_nk} and therefore they should be considered more reliable.

\begin{figure}[t]
\begin{center}
\includegraphics[type=pdf,ext=.pdf,read=.pdf,width=8.5cm]{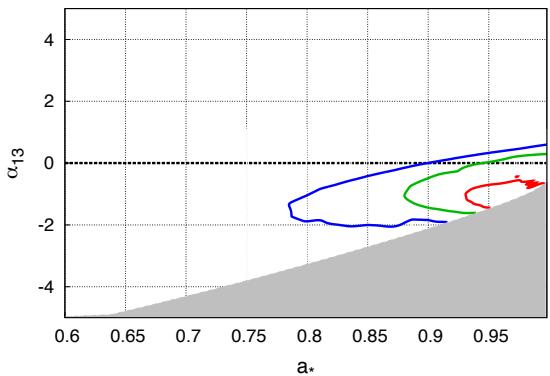}
\end{center}
\vspace{-1.2cm}
\caption{Constraints on the spin parameter $a_*$ and the Johannsen parameter $\alpha_{13}$ from the study in Ref.~\cite{0707} of \textit{NuSTAR} and \textit{Swift} data of 1H0707--495. The red, green, and blue lines indicate, respectively, the 68\%, 90\%, and 99\% confidence level curves for two relevant parameters. The grayed region is ignored because those spacetimes violate the constraint in Eq.~(\ref{eq:boundary-2}). Note that the constraints reported in Ref.~\cite{0707} are slightly different because that study used an earlier version of {\sc relxill\_nk}, and therefore the constraints reported here should be more reliable. \label{f-0707b}}
\end{figure}

For the 2014 observations of \textit{NuSTAR} and \textit{Swift}, we can obtain a good fit already with the model
\begin{flushleft}
\hspace{0.5cm} {\sc tbabs} $\times$ {\sc relxill\_nk} .
\end{flushleft}
The constraints on $a_*$ and $\alpha_{13}$ are shown in Fig.~\ref{f-0707b}. The red, green, and blue lines indicate, respectively, the 68\%, 90\%, and 99\% confidence level curves for two relevant parameters and the grayed region is ignored. Even in this case, the constraints in Fig.~\ref{f-0707b} are slightly different from those reported in Fig.~2 in Ref.~\cite{0707}, because they have been obtained with the latest version of {\sc relxill\_nk} and they are thus more reliable.

\begin{figure}[t]
\begin{center}
\includegraphics[type=pdf,ext=.pdf,read=.pdf,width=8.5cm]{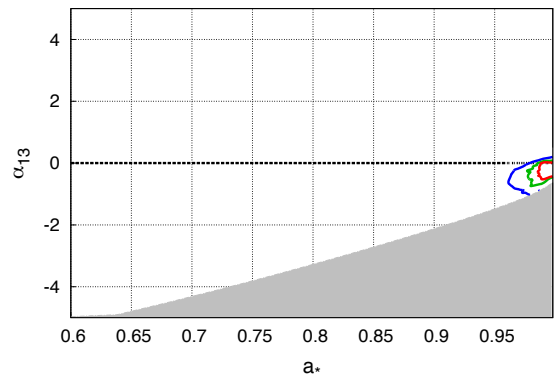}
\end{center}
\vspace{-1.2cm}
\caption{Constraints on the spin parameter $a_*$ and the Johannsen parameter $\alpha_{13}$ from the study in Ref.~\cite{564} of \textit{Suzaku} data of Ark~564. The red, green, and blue lines indicate, respectively, the 68\%, 90\%, and 99\% confidence level curves for two relevant parameters. The grayed region is ignored because those spacetimes violate the constraint in Eq.~(\ref{eq:boundary-2}). \label{f-564}}
\end{figure}

\subsection{Ark~564}

Ark~564 is another Narrow Line Seyfert~1 galaxy. The spectrum is dominated by the reflection component and, previous studies that assumed the Kerr metric, found that the inner edge of the accretion disk extents to very small radii. Ark~564 was observed by \textit{Suzaku} on 26-28 June 2007 for about 80~ks. In Ref.~\cite{564}, we analyzed this observation with {\sc relxill\_nk}. The data can be fit with the following model
\begin{flushleft}
\hspace{0.5cm} {\sc tbabs} $\times$ ( {\sc relxill\_nk} + {\sc xillver} ).
\end{flushleft}
The constraints on the spin parameter $a_*$ and the Johannsen parameter $\alpha_{13}$ are shown in Fig.~\ref{f-564}. The red, green, and blue lines indicate, respectively, the 68\%, 90\%, and 99\% confidence level curves for two relevant parameters. The grayed region is ignored because those spacetimes violate the constraint in Eq.~(\ref{eq:boundary-2}).

\subsection{GX~339--4}

GX~339--4 is an X-ray binary with a stellar-mass black hole and a stellar companion of mass $M_{\rm c} < 2$~$M_\odot$~\cite{heida}. It is quite an active source and typically has an outburst every 2-3~years, so there are several observations in archive. However, many observations are not suitable for our tests of the Kerr metric because the inner edge of the accretion disk is not at the innermost stable circular orbit but is truncated at some larger radius~\cite{jingyi}.

\begin{figure}[t]
\begin{center}
\includegraphics[type=pdf,ext=.pdf,read=.pdf,width=8.5cm]{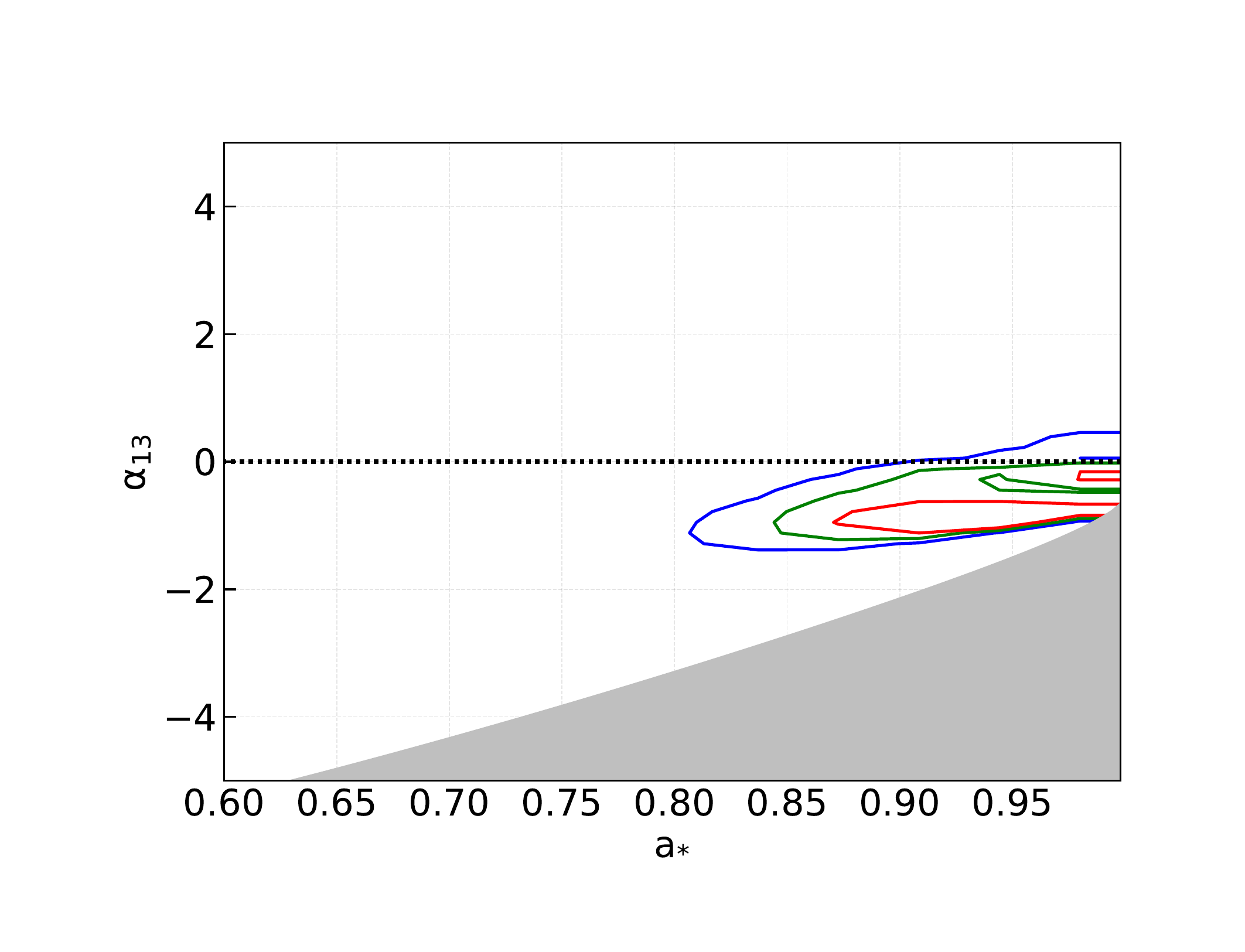}
\end{center}
\vspace{-1.0cm}
\caption{Constraints on the spin parameter $a_*$ and the Johannsen parameter $\alpha_{13}$ from the study in Ref.~\cite{339} of \textit{RXTE} data of GX~339--4. The red, green, and blue lines indicate, respectively, the 1-, 2-, and 3-$\sigma$ confidence level curves for two relevant parameters after MCMC simulations. The grayed region is ignored because those spacetimes violate the constraint in Eq.~(\ref{eq:boundary-2}). \label{f-339}}
\end{figure}

In Ref.~\cite{339}, we analyzed a composite spectrum from the detector CPU-2 of \textit{RXTE} with the highest observed luminosity in the hard state. We reached an unprecedented sensitivity of $\sim 0.1$\% and 40~million counts to capture the faint features in the reflection spectrum. We checked that the data are consistent with the inner edge being at the innermost stable circular orbit. The final model is
\begin{flushleft}
\hspace{0.5cm} {\sc tbabs} $\times$ {\sc gabs} $\times$ ( {\sc relxill\_nk} + {\sc xillver} ).
\end{flushleft}
We performed Markov Chain Monte-Carlo (MCMC) simulations and we obtained the following measurement for the black hole spin $a_*$ and the Johannsen parameter $\alpha_{13}$
\be
a_* = 0.92^{+0.07}_{-0.12} \, , \quad 
\alpha_{13} = -0.76^{+0.78}_{-0.60} \, .
\ee
with a 90\% confidence level for one relevant parameter. The 1-, 2-, and 3-$\sigma$ confidence level curves for two relevant parameters are shown in Fig.~\ref{f-339}.

\begin{figure}[t]
\begin{center}
\includegraphics[type=pdf,ext=.pdf,read=.pdf,width=8.5cm]{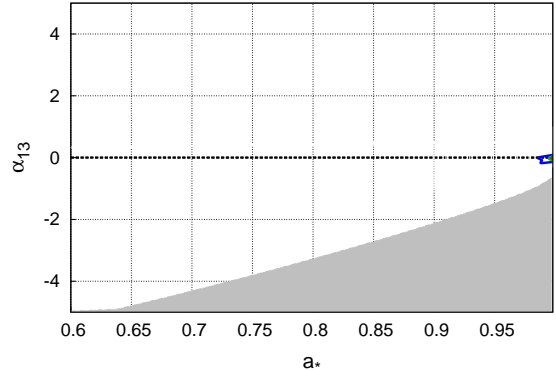}
\end{center}
\vspace{-1.2cm}
\caption{Constraints on the spin parameter $a_*$ and the Johannsen parameter $\alpha_{13}$ from the study in Ref.~\cite{1354} of \textit{NuSTAR} data of GS~1354--645. The red, green, and blue lines indicate, respectively, the 68\%, 90\%, and 99\% confidence level curves for two relevant parameters. The grayed region is ignored because those spacetimes violate the constraint in Eq.~(\ref{eq:boundary-2}). \label{f-1354}}
\end{figure}

\subsection{GS~1354--645}

GS~1354--645 is an X-ray binary with a dynamically confirmed black hole of mass $M \ge 7.6 \pm 0.7$~$M_\odot$ and a companion star of mass $M_{\rm c} \le 1.2$~$M_\odot$~\cite{casares}. The source was discovered in its 1987 outburst by the Japanese X-ray mission \textsl{Ginga}~\cite{makino}. In Ref.~\cite{1354}, we analyzed its July~11 \textsl{NuSTAR} observation of 30~ks. The data can be fit with a simple reflection model
\begin{flushleft}
\hspace{0.5cm} {\sc tbabs} $\times$ {\sc relxill\_nk} .
\end{flushleft}
The constraints on the spin parameter $a_*$ and the Johannsen parameter $\alpha_{13}$ are shown in Fig.~\ref{f-1354}. For a quick comparison with the constraints from the other sources, the plot has the same spin and deformation parameter range as the other plots (the interested reader can find a more detailed plot of the constraints of this source in~\cite{1354}). The constraint on the deformation parameter of this source is so strong here because the data suggest that the inner edge of the accretion disk is extremely close to the compact object. However, this result should be taken with some caution, as well as those from the other sources. We are only taking into account the statistical uncertainties while there are a number of simplifications in our model that can inevitably introduce systematic errors in the final measurement. For example, the Eddington scaled luminosity of this source in the observation analyzed is $L/L_{\rm Edd} \le 0.53$, which is consistent with the $L/L_{\rm Edd} = 0.05$-0.20 range of validity for the standard thin disk model, but it also allows for higher and lower luminosities where our disk model would not be adequate.

\section{Concluding remarks}

Einstein's theory of general relativity has been primarily tested in the weak gravitational regime, while the strong gravity regime is largely unexplored. However, there are many alternative theories of gravity that have the same predictions as Einstein's gravity in weak fields and present deviations when gravity becomes strong. The possibility of testing general relativity in the strong field regime is thus a major goal in modern physics.

Astrophysical black holes offer a quite unique possibility of testing strong gravitational fields. According to general relativity, the spacetime metric around these objects should be well approximated by the Kerr solution. In Ref.~\cite{relxillnk}, we extended the relativistic reflection model {\sc relxill} to test the Kerr black hole hypothesis from the study of the reflection spectrum of thin accretion disks. The new model is called {\sc relxill\_nk}. As of now, we have employed {\sc relxill\_nk} to analyze some X-ray data of the black holes in 1H0707--495, Ark~564, GX~339--4, and GS~1354--645. Current constraints on the spin parameter $a_*$ and the Johannsen deformation parameter $\alpha_{13}$ are summarized in Figs.~\ref{f-0707a}, \ref{f-0707b}, \ref{f-564}, \ref{f-339}, and \ref{f-1354}.

In the future, we plan to analyze the X-ray data of other sources, combine several observations of the same source to perform multi-epoch studies, improve our model {\sc relxill\_nk} in order to reduce the uncertainties of the model, and study a variety of deformations from the Kerr background. If we are able to construct a sufficiently sophisticated model to minimize the systematic effects, the next generation of X-ray mission may start a new era of precise tests of general relativity in the strong field regime with electromagnetic radiation.


\begin{acknowledgments}
This work was supported by the National Natural Science Foundation of China (NSFC), Grant No.~U1531117, and Fudan University, Grant No.~IDH1512060. C.B. also acknowledges support from the Alexander von Humboldt Foundation. A.B.A. also acknowledges the support from the Shanghai Government Scholarship (SGS). S.N. acknowledges support from the Excellence Initiative at Eberhard-Karls Universit\"at T\"ubingen. A.T. also acknowledges support from the China Scholarship Council (CSC), Grant No.~2016GXZR89. 
\end{acknowledgments}


\end{document}